\documentclass[twocolumn,superscriptaddress,floats,prd,nofootinbib]{revtex4-1}
\usepackage{amssymb,amsmath,amssymb,amsfonts,amsthm,stmaryrd,mathrsfs,physics,bbm,mfirstuc}
\usepackage{graphicx}
\usepackage[T1]{fontenc}
\usepackage{enumerate} % advanced enumerate environment
\usepackage{color} % for colored text
\usepackage{graphpap} % for numbered coordinate grid
\usepackage[dvipsnames]{xcolor}

\usepackage{hyperref} % for HyperTeX cross-referencing

\newcommand{\heff}{{H_{\rm eff}}}

\newcommand{\et}{ {\mathcal G} }
\newcommand{\BNU}{School of Physics and Astronomy, \mbox{Key Laboratory of Multiscale Spin Physics (Ministry of Education)}, Beijing Normal University, Beijing 100875, China}

\newcommand{\mysection}[1]{\section{#1}}

\newcounter{mylabelno}

\newcounter{appendix}

\begin{document}

\title{Covariant Hamiltonian Gravity in Spherical Symmetry}

\author{Cong Zhang}
\email{cong.zhang@bnu.edu.cn}
\author{Zhoujian Cao}
\email{zjcao@bnu.edu.cn}
\affiliation{\BNU}

\begin{abstract}
Building upon the covariant Hamiltonian gravity in spherical symmetry, we construct the effective Einstein equation governing gravitational dynamics. For any prescribed one-parameter family of static, spherically symmetric metrics, this equation can be constructed such that the prescribed family constitutes their vacuum solutions.  As an application, we construct generalized Vaidya solutions to the effective Einstein equation and show that the corresponding energy-momentum tensor takes precisely the standard Vaidya form. The results demonstrate that standard null dust sources can serve as matter source for dynamical black-hole geometries beyond general relativity, including regular black holes.  Furthermore, this work suggests that quantum gravity corrections alone may not account for Hawking-like evaporation, as mass loss in the ingoing Vaidya solution necessarily requires a violation of the null energy condition within our effective framework.
\end{abstract} 

\maketitle

%\tableofcontents
\mysection{Introduction}
Understanding quantum gravity remains one of the central challenges in modern theoretical physics. Over the past decades, various approaches have been developed to address this problem, including string theory \cite{polchinski1998string,Aharony:1999ti,Tong:2009np}, loop quantum gravity \cite{Rovelli:1997yv,Thiemann:2001gmi,Rovelli:2004tv,Ashtekar:2004eh,Han:2005km}, asymptotic safety \cite{Niedermaier:2006wt,Ferrero:2025efd,Eichhorn:2026uqj}, causal dynamical triangulations \cite{Ambjorn:2012jv,Ambjorn:2026prt}. Despite considerable progress, a complete theory of quantum gravity remains elusive. Complementary to these fundamental approaches, a pragmatic strategy is to construct an effective description of quantum gravity \cite{Burgess:2003jk,Myers:2010ru,Donoghue:2022eay,Bojowald:2023xat,dePaulaNetto:2023cjw}.  In such descriptions, spacetime is still described by a metric, while quantum gravity effects are incorporated through modifications of the classical gravitational dynamics. Effective models can, in principle, be understood as semiclassical approximations to underlying quantum gravity theories. A notable example is loop quantum cosmology, where the effective dynamics derived from the underlying quantum theory accurately describes the evolution of sharply peaked semiclassical states \cite{Bojowald:2001xe,Ashtekar:2006rx,Ashtekar:2006wn,Taveras:2008ke,Yang:2009fp,Ashtekar:2011ni,MenaMarugan:2011va,Dapor:2017rwv,Assanioussi:2018hee,Zhang:2020mld}.

Beyond cosmology, effective approaches have also been extensively employed to investigate quantum corrections to black-hole spacetimes \cite{Gambini:2013hna,Ashtekar:2018cay,Ashtekar:2018lag,Munch:2021oqn,Kelly:2020lec,Kelly:2020uwj,Husain:2021ojz,Giesel:2022rxi,Lewandowski:2022zce,AlonsoBardaji:2023bww,Giesel:2023tsj,Bojowald:2023djr,Ashtekar:2023cod,Bonanno:2023rzk,ElizagaNavascues:2023gow,Bueno:2024dgm,Giesel:2024mps,Zhang:2024khj,Zhang:2024ney,Zhang:2025ccx,Yang:2025ufs,Sahlmann:2025fde}. However, unlike in loop quantum cosmology, a systematic derivation of effective dynamics from an underlying quantum theory remains unavailable for most existing black-hole models, despite some progress in this direction \cite{Zhang:2020qxw,Zhang:2021xoa,Zhang:2021wex}. In the absence of such a derivation, it is therefore important to identify fundamental principles that can guide and constrain the construction of consistent effective dynamics.  In our previous works \cite{Zhang:2024khj,Zhang:2024ney,Zhang:2025ccx,Yang:2025ufs}, we addressed this issue by developing a systematic framework, which we refer to as covariant Hamiltonian gravity in spherical symmetry, for constructing spherically symmetric effective gravitational dynamics based on the principle of general covariance. Indeed, general covariance requires particular care in the Hamiltonian formulation, where spacetime is foliated into a family of spatial hypersurfaces and general covariance is no longer manifest \cite{TEITELBOIM1973542,Hojman:1976vp,Gambini:2022dec,Bojowald:2022zog,Gambini:2022svq,Han:2022rsx}. Requiring general covariance in this formulation therefore imposes nontrivial restrictions on the admissible forms of the Hamiltonian constraint.

The Hamiltonian formulation, however, presents certain practical limitations when studying matter coupling. In general relativity, matter influences the spacetime metric only through its energy-momentum tensor, so different matter models with the same energy-momentum tensor act as identical sources of gravity. In the Hamiltonian formulation, matter enters through its Hamiltonian and diffeomorphism constraints, whose explicit forms generally depend on the details of the matter fields. A formulation of effective gravitational dynamics directly in terms of the spacetime metric and the energy-momentum tensor would be highly desirable.

Another limitation of the standard Hamiltonian formulation is that it relies on a foliation of spacetime by spacelike hypersurfaces, making it unsuitable for directly working with null foliations. This is particularly inconvenient when studying dynamical spacetimes such as the Vaidya geometry, as well as physical processes such as Hawking radiation, for which null coordinates are particularly useful.

Both limitations can be addressed by constructing an effective Einstein tensor. In the present work, we construct such a tensor from the covariant Hamiltonian gravity in spherical symmetry, allowing the field equations of gravity coupled to matter to be expressed directly in terms of the spacetime metric and the energy-momentum tensor. Moreover, since the resulting equations are tensorial, they can be transformed into arbitrary coordinate systems, including null coordinates, without requiring a null Hamiltonian formulation. 

As an application of the effective Einstein equation, we construct the generalized Vaidya solutions of the effective theory. We find that the corresponding energy-momentum tensor 
retains the standard Vaidya form of classical general relativity. This result demonstrates that null dust can serve as the matter source for dynamical black-hole geometries, including regular black holes. Furthermore, we show that mass loss in the ingoing Vaidya solutions necessarily requires a violation of the null energy condition. This suggests that, within this class of solutions, quantum-gravity corrections to the gravitational dynamics alone may be insufficient to describe Hawking-like evaporation without additional matter effects.

This paper is organized as follows. In Sec.~\ref{sec:II}, we briefly review covariant Hamiltonian gravity in spherical symmetry and the associated reconstruction algorithm. In Sec.~\ref{sec:III}, we construct the effective Einstein tensor and derive its explicit expression in terms of the spacetime metric. In Sec.~\ref{sec:IV}, we consider general matter coupling, introduce the general dynamical metric in the diagonal areal gauge, and evaluate the corresponding effective Einstein tensor. In Sec.~\ref{sec:IV5}, we investigate the generalized Vaidya extension of static, spherically symmetric vacuum geometries. Finally, we summarize our results in Sec.~\ref{sec:conclusion}.

\mysection{Covariant Hamiltonian gravity in spherical symmetry}\label{sec:II}

%Here, the term ``effective'' has two related meanings. First, the spacetime metric is not treated as a fundamental field, but is instead reconstructed from phase-space-dependent functions of the canonical variables. This derived object, which we call the effective metric, need not coincide with its counterpart in classical general relativity. Second, the Hamiltonian constraint governing the dynamics is replaced by an effective one that may differ from its classical expression. 
Motivated by the expectation that such a framework can arise as a semiclassical approximation to an underlying quantum theory of spherically symmetric general relativity, we assume that the canonical phase-space structure retains its classical form. Therefore, in this framework, gravity is described by two canonical pairs, $(E^I,K_I)$ with $I=1,2$, satisfying the Poisson brackets
\begin{equation}
\begin{aligned}
\{K_1(x),E^1(y)\}&=2\delta(x,y),\\
\{K_2(x),E^2(y)\}&=\delta(x,y),
\end{aligned}
\end{equation}
in geometrized units $G=c=1$ \cite{Gambini:2013hna,Zhang:2021xoa}. 

As in classical general relativity, the dynamics is governed by a set of constraints: the diffeomorphism constraint $H_x$ and the effective Hamiltonian constraint $H_{\rm eff}$. The former is assumed to retain its classical expression, as it generates spatial diffeomorphisms, whereas the latter, as mentioned above, may differ from its classical counterpart. Explicitly, the diffeomorphism constraint is given by
\begin{equation}\label{eq:hx}
H_x=E^2\partial_xK_2-\frac{1}{2}K_1\partial_xE^1.
\end{equation}
The form of $H_{\rm eff}$ remains undetermined and will be fixed by imposing further requirements.

Following the framework developed in \cite{Zhang:2024khj,Zhang:2024ney}, the first requirement concerns the constraint algebra. To ensure the consistency of the dynamics, the constraints must form a first-class algebra. Since the diffeomorphism constraint retains its classical form \eqref{eq:hx}, the Poisson brackets involving $H_x$ are already fixed. It therefore remains to specify the bracket between two effective Hamiltonian constraints, which we require to take the form
\begin{equation}\label{eq:poisson}
\{\heff[N],\heff[M]\}
=H_x\left[S\left(N\partial_x M-M\partial_x N\right)\right],
\end{equation}
where $S=\mu E^1/(E^2)^2$ is the structure function and $G[f]\equiv\int Gf\,\dd x$ denotes the smeared constraint. The deformation factor $\mu$ captures possible quantum-gravity corrections. It equals unity in classical general relativity but may depend on the phase-space variables in the effective theory.

The second requirement concerns general covariance, which is not guaranteed by the closure of the constraint algebra alone \cite{Bojowald:2023xat}. The subtlety arises from the Hamiltonian formulation, in which four-dimensional spacetime is foliated into a family of three-dimensional spatial hypersurfaces, thereby obscuring the manifest covariance of the theory. To solve the equations of motion in the Hamiltonian framework, one must specify the Lagrange multipliers, namely the lapse function $N$ and the shift vector $N^x$. General covariance then requires that different choices of $N$ and $N^x$, starting from the same physical initial data, yield spacetime metrics related by spacetime diffeomorphisms. To formulate this requirement explicitly, we must first specify how the spacetime metric is constructed from the canonical variables, lapse function, and shift vector. Motivated by the geometric interpretation of the constraint algebra \eqref{eq:poisson} in terms of hypersurface deformations \cite{Bojowald:2023xat,Zhang:2024khj,Zhang:2024ney}, the effective spacetime metric reads
\begin{equation}\label{eq:effmetric}
\dd s^2=-N^2\dd t^2+\frac{(E^2)^2}{\mu E^1}(\dd x+N^x\dd t)^2+E^1\dd\Omega^2.
\end{equation}
The requirement of general covariance, namely that different choices of $N$ and $N^x$ yield spacetime metrics related by diffeomorphisms, imposes further restrictions on the form of $\heff$. To express these restrictions explicitly, it is convenient to introduce the scalars
\begin{equation*}
\begin{aligned}
&s_1=E^1,\quad
s_2=K_2,\quad
s_3=\frac{K_1}{E^2},\\
&s_4=\frac{\partial_x s_1}{E^2},\quad
s_5=\frac{\partial_x s_4}{E^2}.
\end{aligned}
\end{equation*}
In terms of these variables, the generally covariant effective Hamiltonian constraint takes the form \cite{Zhang:2024khj,Zhang:2024ney,Zhang:2025ccx}
\begin{equation}\label{eq:HinMeff}
\heff=-2E^2\Big[
\partial_{s_1}M_{\rm eff}
+\frac{\partial_{s_2}M_{\rm eff}}{2}s_3
+\frac{\partial_{s_4}M_{\rm eff}}{s_4}s_5
+\mathcal R
\Big],
\end{equation}
where $\mathcal R$ is a function of $s_1$ and $M_{\rm eff}$, while $M_{\rm eff}$ is a function of $s_1$, $s_2$, and $s_4$, implicitly defined by
\begin{equation}\label{eq:determinMeff}
\big(\partial_{s_2}M_{\rm eff}\big)^2
=\frac{\mu s_1}{4}\big(s_4^2+\mathcal Z\big),
\end{equation}
for some function $\mathcal Z(s_1,M_{\rm eff})$. Moreover, general covariance further requires $\mu$ to be a function of $s_1$ and $M_{\rm eff}$. Thus, different models in covariant Hamiltonian gravity in spherical symmetry are characterized by different choices of the three functions $\mathcal R$, $\mu$, and $\mathcal Z$. 

For a given choice of the three functions $\mathcal R$, $\mu$, and $\mathcal Z$, the resulting theory admits a unique one-parameter family of static, spherically symmetric vacuum solutions, as shown in \cite{Zhang:2025ccx}. In Schwarzschild-like coordinates, the metric takes the form
\begin{equation}\label{eq:dsmetric}
\dd s^2=\frac{1}{4}N_o^2\mathcal Z\dd t^2-\frac{4}{\mu\mathcal Z}\dd x^2+x^2\dd\Omega^2,
\end{equation}
where $\dd\Omega^2=\dd\theta^2+\sin^2\theta\,\dd\phi^2$ is the line element on $\mathbb S^2$, and
\begin{equation}\label{eq:No}
N_o=\exp\left(2\int x\partial_{M_{\rm eff}}\mathcal R\,\dd x\right).
\end{equation}
Here, the functions $\partial_{M_{\rm eff}}\mathcal R$, $\mathcal Z$, and $\mu$ are evaluated at $s_1=x^2$ and $$M_{\rm eff}=M(x;m),$$ where $M(x;m)$ satisfies
\begin{equation}\label{eq:RinM}
\partial_xM+2x\mathcal R(x^2,M)=0,
\end{equation}
with $m$ denoting the integration constant. This equation follows from the constraints $H_{\rm eff}=0=H_x$. Moreover, as shown in \cite{Zhang:2025ccx}, $\partial_t M_{\rm eff}=\{M_{\rm eff},H_{\rm eff}[N]+H_x[N^x]\}\approx 0$, which ensures that $m$ is also independent of time. Here, $\approx$ denotes equality on the constraint surface defined by $H_{\rm eff}=0=H_x$.

It is worth emphasizing the distinction between $M_{\rm eff}$, $M(x;m)$, and $m$. The function $M_{\rm eff}(s_1,s_2,s_4)$ is defined on the gravitational phase space and enters the construction of the effective Hamiltonian constraint. The explanation following Eq.~\eqref{eq:RinM} implies that, when evaluated on the vacuum solutions, $M_{\rm eff}$ becomes $M(x;m)$, where $m$ is a constant labeling the one-parameter family of vacuum solutions.

\subsection{Restricting the form of $\mu$, $\mathcal Z$ and $\mathcal R$}
The functions $\mu$, $\mathcal Z$, and $\mathcal R$ can be further constrained by requiring that a prescribed metric be a vacuum solution of the theory. To this end, we consider an arbitrary static, spherically symmetric metric depending on a mass parameter $m$, expressed in Schwarzschild-like coordinates as
\begin{equation}\label{eq:metric00}
\dd s^2=-F(x;m)\dd t^2+\frac{\dd x^2}{H(x;m)}+x^2\dd\Omega^2.
\end{equation}
 
 Comparing Eqs.~\eqref{eq:dsmetric} and \eqref{eq:metric00}, we obtain
\begin{equation}\label{eq:condition}
\frac{1}{4}N_o^2\mathcal Z=-F,\qquad
\frac{\mu\mathcal Z}{4}=-H,
\end{equation}
as the conditions for the theory to admit the metric \eqref{eq:metric00} as a vacuum solution. Here, the substitutions $E^1=x^2$ and $M_{\rm eff}=M(x;m)$ must be made on the left-hand sides, so that both sides of the equations are expressed as functions of $x$ and $m$. Equation~\eqref{eq:condition} imposes constraints on the three functions $\mu$, $\mathcal Z$, and $\mathcal R$. However, it provides only two equations for these three unknown functions. Consequently, one functional degree of freedom remains undetermined. In our previous work \cite{Zhang:2025ccx}, this remaining functional freedom was parametrized by the function $\mathcal R$. In the present work, we adopt an alternative parametrization in terms of $M(x;m)$, the on-shell value of the effective mass function $M_{\rm eff}$.

Equation~\eqref{eq:RinM} has related $\mathcal R$ to the mass function $M(x;m)$. To further express the function $\mathcal Z$ and $\mu$ in terms of $M(x;m)$, let us consider the derivative of   \eqref{eq:RinM} with respect to $m$, leading to
\begin{equation}\label{eq:eq4dmM}
\partial_x\partial_mM+2x(\partial_{M_{\rm eff}}\mathcal R)\partial_mM=0.
\end{equation}
In addition, by the expression of $N_o$ in \eqref{eq:No}, it also satisfies the equation 
\begin{equation}\label{eq:equation4Noinverse}
\partial_x\left(\frac{1}{N_o}\right)+2x(\partial_{M_{\rm eff}}\mathcal R)\frac{1}{N_o}=0.
\end{equation}
Combining \eqref{eq:equation4Noinverse} with \eqref{eq:eq4dmM}, we find that 
\begin{equation}\label{eq:CmindmMNo}
\partial_mM N_o=C(m)
\end{equation}
for some function $C(m)$. We now show that $C(m)$ can be set to $1$ by an appropriate choice of normalization. Indeed, the expression for $N_o$ in Eq.~\eqref{eq:No} is defined only up to a multiplicative factor independent of $x$, which may depend on $m$. Equivalently, we may write
\begin{equation}
N_o=A(m)\exp\left(2\int_{x_o}^{x}\xi\partial_{M_{\rm eff}}\mathcal R,\dd\xi\right),
\end{equation}
where $A(m)$ represents the normalization freedom. Provided $\partial_mM\neq 0$, this freedom allows us to choose $C(m)=1$, so that
\begin{equation}\label{eq:NoindmM}
N_o=\frac{1}{\partial_mM}.
\end{equation}
Substituting this result into Eq.~\eqref{eq:condition} and using Eq.~\eqref{eq:RinM}, we obtain
\begin{equation}\label{eq:reconstructionEqs}
\begin{aligned}
\mathcal R&=-\frac{\partial_x M}{2x},\quad
\mu=\frac{H}{F(\partial_m M)^2},\\
\mathcal Z&=-4F(\partial_mM)^2.
\end{aligned}
\end{equation}
These equations express $\mathcal R$, $\mu$, and $\mathcal Z$ as functions of $x$ and $m$, which can then be reexpressed as functions of $(s_1,M_{\rm eff})$ through the relations $s_1=x^2$ and $M_{\rm eff}=M(x;m)$.

\section{The effective Einstein tensor}\label{sec:III}

We now turn to the spacetime-tensor formulation of the effective gravitational equations. We will consider a general matter coupling.  As shown in \cite{Yang:2025ufs}, when matter couples only to the metric and not to its derivatives, the coupled system remains generally covariant if the classical metric in the matter action is replaced by the effective metric \eqref{eq:effmetric}. Accordingly, in this work, we restrict our consideration to matter action $S_m$ depending only on $g_{\mu\nu}$, matter fields and the derivatives of the matter fields.  For the gravity field, applying the inverse Legendre transformation yields
\begin{equation}
S_g=-\int\dd t\dd x\left(\frac{1}{2}K_1\dot E^1+K_2\dot E^2+N\heff+N^xH_x\right).
\end{equation}
Its variation reads 
\begin{equation}\label{eq:deltaSg}
\begin{aligned}
\delta S_g=&\int\dd t\dd x\Bigg[\frac{1}{2}\left(-\dot E^1+\{E^1,  H_{\rm eff}[ N]+H_x[N^x]\}\right)\delta K_1\\
&+\left(-\dot E^2+\{E^2, H_{\rm eff}[ N]+H_x[N^x]\}\right)\delta K_2\\
&+\frac{1}{2}\left(\dot K_1-\{K_1,   H_{\rm eff}[ N]+H_x[N^x]\}\right)\delta E^1\\
&+ \left(\dot K_2-\{K_2, H_{\rm eff}[ N]+H_x[N^x]\}\right)\delta E^2\\
&- H_{\rm eff}\delta N-H_x\delta N^x\Bigg. 
\end{aligned}
\end{equation}

Although the following calculations can be carried out directly in terms of the effective metric $g_{\mu\nu}$, it is convenient to introduce an auxiliary two-dimensional metric $\mathring g_{ab}$. This metric is not an independent dynamical field, but is constructed from the canonical variables and the deformation factor $\mu$. Its introduction simplifies the subsequent calculations. Specifically, for $a,b=0,1$, we define
\begin{equation}
\mathring g_{ab}\dd x^a\dd x^b=-\mathring N^2\dd t^2+\frac{(E^2)^2}{E^1}(\dd x+N^x\dd t)^2,
\end{equation}
with 
\begin{equation}
\mathring N=\sqrt{\mu} N. 
\end{equation}
The 4-dimensional metric can be written in terms of $\mathring g_{ab}$ as 
\begin{equation}
g_{\mu\nu}\dd x^\mu\dd x^\nu=\frac{1}{\mu}\mathring g_{ab}\dd x^a\dd x^b+E^1\dd\Omega^2. 
\end{equation}
Since $\mu$ and $E^1$ are both scalars, $\mathring g_{ab}$ defines a tensor in the theory which are covariant with respect to the effective metric $g_{\mu\nu}$. In terms of $\mathring g_{ab}$, we can rewrite the matter action $S_m$ as 
\begin{equation}
S_m=S_m[\mathring g_{ab},\mu,E^1,\psi,\cdots]
\end{equation}
where $\psi$ denotes the matter fields and $\cdots$ represents the derivatives of $\psi$. We then define $\mathring T_{ab}$, $\mathcal E^{(\mu)}$, and $T^{(\theta)}$ through the variation of the matter action:
\begin{equation}\label{eq:deltasm}
\begin{aligned}
\delta S_m=-\int \dd t\dd x\sqrt{-\mathring g}&\Bigg(\mathring T_{ab}\delta\mathring g^{ab}+\mathcal E^{(\mu)}\delta\mu\\
&+T^{(\theta)}\delta E^1+\cdots\Bigg),
\end{aligned}
\end{equation}
where $\cdots$ denotes the terms arising from variations of the matter fields $\psi$. Here, $\mathring T_{ab}$, $\mathcal E^{(\mu)}$, and $T^{(\theta)}$ are related to the four-dimensional energy-momentum tensor $T_{\mu\nu}$ through
\begin{equation}
\begin{aligned}
&\frac12\int \dd^4x\sqrt{-g}\,
T_{\mu\nu}\delta g^{\mu\nu}\\
=&\int \dd t\dd x\sqrt{-\mathring g}
\left(
\mathring T_{ab}\delta\mathring g^{ab}
+\mathcal E^{(\mu)}\delta\mu
+T^{(\theta)}\delta E^1
\right),
\end{aligned}
\end{equation}
which yields
\begin{equation}\label{eq:TfromTo}
\begin{aligned}
T_{ab}
&=\frac{\mathring T_{ab}}{2\pi E^1},\ T_{\theta\theta}
=-\frac{\mu E^1}{4\pi}T^{(\theta)}=\frac{T_{\phi\phi}}{\sin^2\theta},\\
\mathcal E^{(\mu)}
&=\frac{1}{\mu}\mathring g^{ab}\mathring T_{ab}.
\end{aligned}
\end{equation}
These relations show that, for matter actions depending on the physical metric $g_{\mu\nu}$ but not on any additional independent coupling to $\mu$, the quantity $\mathcal E^{(\mu)}$ is not independent. Instead, it is completely determined by the trace of the two-dimensional energy-momentum tensor $\mathring T_{ab}$.

By definition, $\mathring g^{ab}$ reads 
\begin{equation}
g^{ab}\partial_a\partial_b=-\frac{1}{\mathring N^2}\mathfrak N \otimes \mathfrak N+\frac{E^1}{(E^2)^2}\partial_x\otimes\partial_x,
\end{equation}
with
\begin{equation*}
\mathfrak N=\partial_t-N^x\partial_x.
\end{equation*}
Then, the variation of $\mathring g^{ab}$ can be calculated as
\begin{equation}
\begin{aligned}
\delta \mathring g^{ab}=&\frac{2}{\mathring N^3}\left(\sqrt{\mu}\delta N+\frac{\mathring N}{2\mu}\delta \mu \right)\mathfrak N^a\mathfrak N^b+\frac{2}{\mathring N^2}\mathfrak N^{(a}\delta_1^{b)}\delta N^x\\
&+\left(\frac{1}{(E^2)^2}\delta E^1-\frac{2E^1}{(E^2)^3}\delta E^2\right)\delta^a_1\delta^b_1.
\end{aligned}
\end{equation}
Substituting this result into \eqref{eq:deltasm}, we get 
\begin{equation}\label{eq:deltaSm1}
\begin{aligned}
\delta S_m=&-\int \dd t\dd x\frac{\sqrt{\mu}N E^2}{\sqrt{E^1}}\Bigg[\frac{2}{\mu N^3}\mathring T_{\mathfrak N\mathfrak N} \delta N+\frac{2}{\mu N^2} \mathring T_{ \mathfrak N x}\delta N^x\\
&-\frac{2E^1}{(E^2)^3}\mathring T_{xx}\delta E^2+\left(\frac{\mathring T_{xx}}{(E^2)^2}+T^{(\theta)}\right)\delta E^1\\
&+\left(\mathcal E^{(\mu)}+\frac{\mathring T_{\mathfrak N\mathfrak N}}{\mu^2 N^2 }\right)\delta\mu+\cdots\Bigg]
\end{aligned}
\end{equation}
with
\begin{equation*}
\begin{aligned}
\mathring T_{\mathfrak N\mathfrak N}&=\mathring T_{ab}\mathfrak N^a\mathfrak N^b,\quad \mathring T_{\mathfrak Nx}=\mathring T_{ab}\mathfrak N^a\delta^b_1,\\ 
\mathring T_{xx}&=\mathring T_{ab}\delta^a_1\delta^b_1.
\end{aligned}
\end{equation*}
%Since $\mu$ depends on $s_1$ and $M_{\rm eff}$, the last term in Eq.~\eqref{eq:deltaSm1} can be simplified to
%\begin{equation}
%\begin{aligned}
%&\int \dd t\dd x\sqrt{-\mathring g} \left(\mathcal E^{(\mu)}+\frac{\mathring T_{\mathfrak N\mathfrak N}}{\mu^2 N^2 }\right)\delta\mu\\
%=&\int \dd t\dd x\sqrt{-\mathring g} \Bigg\{\left(\mathcal E^{(\mu)}+\frac{\mathring T_{\mathfrak N\mathfrak N}}{\mu^2 N^2 }\right)\left(\mu_1\delta E^1+\right)\\
%&\left(\mathcal E^{(\mu)}+\frac{\mathring T_{\mathfrak N\mathfrak N}}{\mu^2 N^2 }\right)\left(\mu_2\delta K_2-\mu_4\frac{\partial_xE^1}{(E^2)^2}\delta E^1\right) \Bigg]
%\end{aligned}
%\end{equation}
%where we define
%\begin{equation}
%\begin{aligned}
%&\mu_1=\partial_{s_1}\mu+\partial_{M_{\rm eff}}\mu\partial_{s_1}M_{\rm eff},\quad \mu_2=\partial_{M_{\rm eff}}\mu\partial_{s_2}M_{\rm eff}\\
% &\mu_4=\partial_{M_{\rm eff}}\mu\partial_{s_4}M_{\rm eff}
%\end{aligned}
%\end{equation}

Combining this result \eqref{eq:deltaSm1} with Eq.~\eqref{eq:deltaSg} and imposing the stationary-action principle $\delta S_g+\delta S_m=0$, we obtain:
\begin{itemize}
\item[(1)] Varying $N$ while keeping all other variables fixed yields 
\begin{equation}\label{eq:Tnn}
 \mathring T_{ \mathfrak N \mathfrak N}=-\frac{\sqrt{\mu E^1}N^2}{2E^2}H_{\rm eff}.
\end{equation}
\item[(2)] Varying  $N^x$ while keeping all other variables fixed leads to
\begin{equation}\label{eq:Tnx}
\mathring T_{ \mathfrak Nx}=-\frac{ N\sqrt{\mu E^1}} {2E^2}H_x
\end{equation}

\item[(3)]Varying $E^2$ while keeping all other variables fixed, and noting that $\delta\mu=\partial_{E^2}\mu\,\delta E^2$ in this case, we obtain
\begin{equation}\label{eq:Txx}
\begin{aligned}
&\frac{2N\sqrt{\mu E^1}}{(E^2)^2}\mathring T_{xx}+ \frac{\partial\mu}{\partial E^2}\frac{N}{2\mu}\left(H_{\rm eff}-\frac{2\sqrt{\mu}^3E^2}{\sqrt{E^1}}\mathcal E^{(\mu)}\right)\\
=&-\partial_t K_2+\{K_2, H_{\rm eff}[ N]+H_x[N^x]\},
\end{aligned}
\end{equation}
where we used \eqref{eq:Tnn} to replace $\mathring T_{\mathfrak N\mathfrak N}$ by $\heff$. To simplify the formulation, we observe that 
\begin{equation}\label{eq:K2tildeN}
\begin{aligned}
&\{K_2(x), H_{\rm eff}[ N]\}\\
=&\left\{K_2(x),\int\dd y\frac{H_{\rm eff}(y)}{\sqrt{\mu(y)}}\lambda(y)\right\}\Bigg|_{\lambda=\mathring N}\\
&+H_{\rm eff}\frac{N}{2\mu}\frac{\partial \mu}{\partial E^2}
\end{aligned}
\end{equation}
with which, we can simplify  \eqref{eq:Txx} to be
\begin{equation}\label{eq:Txx1}
\begin{aligned}
\mathring T_{xx}+\frac{\mu_4 E^2\partial_xE^1}{2E^1}\mathcal E^{(\mu)}=\frac{(E^2)^2}{2N\sqrt{\mu E^1}}  \mathring{\mathscr D}K_2
\end{aligned}
\end{equation}
where we have introduced  the operator $\mathring{\mathscr D}$ as
\begin{equation}
\begin{aligned}
\mathring{\mathscr D}=-\partial_t+\{\,\cdot\,,\mathring H_{\rm eff}[\lambda]+H_x[N^x]\}\big|_{\lambda=\mathring N},
\end{aligned}
\end{equation}
with 
\begin{equation}
\mathring H_{\rm eff}=\frac{H_{\rm eff}}{\sqrt{\mu}}.
\end{equation}
The notation $\big|_{\lambda=\mathring N}$ indicates that the Poisson bracket is evaluated first, treating $\lambda$ as a phase-space-independent smearing function, and only then is $\lambda$ replaced by $\mathring N$. Moreover, in Eq. \eqref{eq:Txx1}, we used 
\begin{equation}
\frac{\partial\mu}{\partial E^2} =-\mu_4\frac{\partial_xE^1}{(E^2)^2},\quad \mu_4=\partial_{M_{\rm eff}}\mu\partial_{s_4}M_{\rm eff}.
\end{equation}

\item[(4)]Varying $E^1$ while keeping all other variables fixed, and noting that $\mu$ depends on $\partial_xE^1$ and $E^1$, we get
\begin{equation}\label{eq:Txx2}
\begin{aligned}
&\frac{E^2}{2E^1}  \mathring{\mathscr D}K_2-  \frac{\sqrt{\mu}N  \mu_4 \partial_xE^1}{2\sqrt{E^1}^3}\mathcal E^{(\mu)}+\frac{\sqrt{\mu}N E^2}{\sqrt{E^1}}T^{(\theta)}\\
&+\left(\frac{\sqrt{\mu}N E^2 \mathcal E^{(\mu)}}{\sqrt{E^1}}-\frac{N}{2\mu}H_{\rm eff}\right)\frac{\partial\mu}{\partial E^1}\\
&-\partial_x\left[\left(\frac{\sqrt{\mu}N E^2 \mathcal E^{(\mu)}}{\sqrt{E^1}}-\frac{N}{2\mu}H_{\rm eff}\right)\frac{\partial\mu}{\partial(\partial_xE^1)}\right]\\
=&\frac{1}{2}\left(\partial_t K_1-\{K_1,   H_{\rm eff}[ N]+H_x[N^x]\}\right)
\end{aligned}
\end{equation}
where  \eqref{eq:Tnn} and \eqref{eq:Txx} are applied. Similarly as Eq. \eqref{eq:K2tildeN}, the Poisson bracket between $K_1$ and $\heff[N]$ can be rewritten as 
\begin{equation}\label{eq:K1Ho}
\begin{aligned}
&\int\dd t\dd x\{K_1,H_{\rm eff}[N]\}\delta E^1\\
=&\int\dd t\dd x\frac{N}{\mu}  H_{\rm eff}\left(\frac{\partial \mu}{\partial E^1}\delta E^1+ \frac{\partial \mu}{\partial \partial_xE^1}\partial_x\delta E^1\right)+\\
&\int\dd t\dd x\{K_1,\mathring H_{\rm eff}[\lambda]\}\Big|_{\lambda=\mathring N}\delta E^1.
\end{aligned}
\end{equation}
With this result, Eq. \eqref{eq:Txx2} can be simplified to get
\begin{equation}\label{eq:Txx2p}
\begin{aligned}
&\frac{\sqrt{\mu}N  E^2}{2\sqrt{E^1}^3}\mathcal E^{(\mu)} \left(2E^1\mu_1 -  \frac{ \mu_4 \partial_xE^1}{E^2}\right)\\
&-\partial_x\left(\frac{\sqrt{\mu}N \mu_4}{\sqrt{E^1}}\mathcal E^{(\mu)}\right)+\frac{\sqrt{\mu}N E^2}{\sqrt{E^1}}T^{(\theta)}\\
=&-\frac{E^I \mathring{\mathscr D}K_I}{2E^1} 
\end{aligned}
\end{equation}
where we introduced 
\begin{equation}
\mu_1=\partial_{s_1}\mu+\partial_{M_{\rm eff}}\mu\partial_{s_1}M_{\rm eff}
\end{equation}
and used 
\begin{equation}
\frac{\partial\mu}{\partial(\partial_xE^1)}=\frac{\mu_4}{E^2}
\end{equation}

\item[(5)] Varying $K_1$ while keeping all other variables fixed yields
\begin{equation}
\delta \mathring g^{ab}=0,
\end{equation} 
leading to
\begin{equation}\label{eq:eqE1}
\begin{aligned}
\mathring{\mathscr D}E^1=0.
\end{aligned}
\end{equation}

\item[(6)] Varying $K_2$ while keeping all other variables fixed yields, and noting that $\mu$ depends on $K_2$, we get
\begin{equation}\label{eq:finalEq}
\begin{aligned}
&\left(\frac{\sqrt{\mu}N E^2}{\sqrt{E^1}}\mathcal E^{(\mu)}-\frac{N}{2\mu}H_{\rm eff}\right)\frac{\partial \mu}{\partial K_2}\\
=&-\partial_t E^2+\{E^2, H_{\rm eff}[ N]+H_x[N^x]\}
\end{aligned}
\end{equation}
where \eqref{eq:Tnn} is applied. Similar as Eq. \eqref{eq:K2tildeN}, the Poisson bracket between $E^2$ and $\heff[N]$ can be simplified to
\begin{equation}\label{eq:E2tildeN}
\begin{aligned}
&\{E^2(x), H_{\rm eff}[ N]\}\\
=&\left\{E^2(x),\int\dd y\frac{H_{\rm eff}(y)}{\sqrt{\mu(y)}}\lambda(y)\right\}\Big|_{\lambda=\mathring N}\\
&-H_{\rm eff}\frac{N}{2\mu}\frac{\partial \mu}{\partial K_2}
\end{aligned}
\end{equation}
Substituting this result into \eqref{eq:finalEq}, we get
\begin{equation}\label{eq:finalEq1}
\begin{aligned}
\frac{\sqrt{\mu} \mu_2N E^2}{\sqrt{E^1}}\mathcal E^{(\mu)}=\mathring{\mathscr D}E^2,
\end{aligned}
\end{equation}
where we introduced 
\begin{equation}\label{eq:definemu2}
\mu_2=\partial_{M_{\rm eff}}\mu\partial_{s_2}M_{\rm eff}. 
\end{equation}
\end{itemize}

The above equations \eqref{eq:Tnn}, \eqref{eq:Tnx}, \eqref{eq:Txx1}, \eqref{eq:Txx2p}, \eqref{eq:eqE1}, and \eqref{eq:finalEq1} follow directly from the variation of the total action, without assuming any Einstein-like field equations. We now reorganize them into a covariant tensorial form by identifying the corresponding gravitational and matter contributions. To this end, we define the tensors $\mathring \et_{\mu\nu}$ and $\mathcal T_{\mu\nu}$ as follows:
\begin{equation}\label{eq:einsteineqs}
\begin{aligned}
&\mathring \et_{\alpha\beta}\dd x^\alpha\dd x^\beta\\
=&-\frac{N^2\sqrt{\mu E^1} H_{\rm eff}}{2E^2}\dd t^2- \frac{N\sqrt{\mu E^1} }{E^2}H_x(\dd x+N^x\dd t)\dd t\\
&+\frac{(E^2)^2}{2 N\sqrt{\mu E^1}}\mathring{\mathscr D}K_2\left(\dd x+N^x\dd t\right)^2+\frac{E^I\mathring{\mathscr D}K_I}{4N\sqrt{\mu E^1}E^2}\dd\Omega^2
\end{aligned}
\end{equation}
and 
\begin{equation}\label{eq:calT}
\begin{aligned}
&\mathcal T_{\alpha\beta}\dd x^\alpha\dd x^\beta\\
=&\mathring T_{ ab}\dd x^a\dd x^b+\frac{\mu_4 E^2\partial_xE^1}{2E^1}\mathcal E^{(\mu)}(\dd x+N^x\dd t)^2\\
&+\Bigg[ -\frac{T^{(\theta)}}{2}+\left( \frac{ \mu_4 \partial_xE^1}{E^2}-2E^1\mu_1 \right)\frac{\mathcal E^{(\mu)}}{4 E^1}\\
&+\frac{\sqrt{E^1}}{2N\sqrt{\mu}E^2}\partial_x\left(\frac{\sqrt{\mu}N \mu_4}{\sqrt{E^1}}\mathcal E^{(\mu)}\right)\Bigg]\dd\Omega^2
\end{aligned}
\end{equation}
 Then, all of equations of motion can be written as
\begin{equation}\label{eq:finaltensor}
\begin{aligned}
\mathring \et_{\alpha\beta}=&\mathcal T_{\alpha\beta},\\
\mathring{\mathscr D}E^1=&0,\\
\mathring{\mathscr D}E^2=&\frac{\sqrt{\mu} \mu_2N E^2}{\sqrt{E^1}}\mathcal E^{(\mu)},
\end{aligned}
\end{equation}
where the last two equations  define $K_I$ in terms of $E^I$, their derivatives and $\mathcal E^{(\mu)}$. In what follows, we will call $\mathring\et_{\mu\nu}$ the effective Einstein tensor.

 \subsection{Explicit expression of $\mathring \et_{ab}$}\label{sec:einsteinnx0}
Now let us calculate the explicit expression of $\mathring \et$. Equation~\eqref{eq:HinMeff} implies the following relation between $\heff$ and $H_x$:
\begin{equation}\label{eq:HeffinHxRMeff}
\heff-2\frac{\partial_{s_2}M_{\rm eff}}{\partial_xE^1} H_x=-\frac{2E^2}{\partial_xE^1}\left(\mathcal R\partial_x E^1+\partial_xM_{\rm eff}\right).
\end{equation}
Taking advantage of this result, we can simplify the last two equations in Eq.~\eqref{eq:finaltensor}, to obtain 
\begin{equation}\label{eq:dtE1}
\begin{aligned}
\mathcal L_{\mathfrak N}E^1=&2N\partial_{s_2}M_{\rm eff}\\
\mathcal L_{\mathfrak N}E^2=&N\mu_2\left(\frac{H_{\rm eff}}{2 \mu}-\frac{\sqrt{\mu}  E^2}{\sqrt{E^1}}\mathcal E^{(\mu)}\right)-\frac{2  N \partial_{s_2}^2M_{\rm eff}}{\partial_xE^1}H_x\\
&+2 N E^2  \partial_{M_{\rm eff}}\mathcal R\partial_{s_2}M_{\rm eff} +\frac{2 N E^2\partial_x\partial_{s_2}M_{\rm eff}}{\partial_xE^1}
\end{aligned}
\end{equation}
where we introduced 
\begin{equation}
\begin{aligned}
\mathcal L_{\mathfrak N}E^1&=\partial_tE^1-N^x\partial_xE^1,\\
\mathcal L_{\mathfrak N}E^2&=\partial_tE^2-\partial_x(N^xE^2)
\end{aligned}
\end{equation}
Solving $\partial_{s_2}M_{\rm eff}$ from the first equation and substituting the result into the second, we obtain
\begin{equation}\label{eq:dtE2}
\begin{aligned}
&\frac{\partial_{M_{\rm eff}}\mu\mathcal L_{\mathfrak N}E^1}{4\mu}H_{\rm eff}-\frac{2  N \partial_{s_2}^2M_{\rm eff}}{\partial_xE^1}H_x\\
=&\mathcal L_{\mathfrak N}E^2+\frac{\sqrt{\mu}  E^2\mathcal L_{\mathfrak N}E^1\partial_{M_{\rm eff}}\mu}{2\sqrt{E^1}}\mathcal E^{(\mu)}\\
&- E^2 \mathcal L_{\mathfrak N}E^1 \partial_{M_{\rm eff}}\mathcal R-\frac{N E^2}{\partial_xE^1}\partial_x\left(\frac{\mathcal L_{\mathfrak N}E^1}{N}\right),
\end{aligned}
\end{equation}
with applying Eq. \eqref{eq:definemu2}.
Equations~\eqref{eq:dtE2} and \eqref{eq:HeffinHxRMeff} form a system of two linear equations for $\heff$ and $H_x$. Solving this system, we obtain
\begin{equation}\label{eq:HeffHx1}
\begin{aligned}
&H_{\rm eff}
=\frac{4\mu}{D\,\partial_xE^1}
\Bigg(4N^2E^2
\left(\mathcal R\partial_xE^1+\partial_xM_{\rm eff}\right)
\partial_{s_2}^2M_{\rm eff}\\
&+\partial_xE^1\mathcal L_{\mathfrak N}E^1\mathcal L_{\mathfrak N}E^2-N E^2\mathcal L_{\mathfrak N}E^1\partial_x\left(\frac{\mathcal L_{\mathfrak N}E^1}{N}\right)\\
&
-E^2\partial_xE^1(\mathcal L_{\mathfrak N}E^1)^2
\partial_{M_{\rm eff}}\mathcal R\\
&+\frac{\sqrt{\mu}E^2\partial_xE^1
(\mathcal L_{\mathfrak N}E^1)^2\partial_{M_{\rm eff}}\mu}
{2\sqrt{E^1}}\mathcal E^{(\mu)}
\Bigg),
\end{aligned}
\end{equation}
and
\begin{equation}\label{eq:HeffHx2}
\begin{aligned}
H_x=&\frac{-2N}{D}\Bigg(
2\mu E^2\mathcal L_{\mathfrak N}E^1\partial_xE^1\partial_{M_{\rm eff}}\mathcal R-2\mu\partial_xE^1\mathcal L_{\mathfrak N}E^2\\
&-\mathcal L_{\mathfrak N}E^1 E^2\left(\mathcal R\partial_xE^1+\partial_xM_{\rm eff}\right)\partial_{M_{\rm eff}}\mu\\
&+2\mu E^2 N
\partial_x\left(\frac{\mathcal L_{\mathfrak N}E^1}{N}\right)\\
&-\frac{\sqrt{\mu}^3\,E^2\partial_xE^1
\mathcal L_{\mathfrak N}E^1\partial_{M_{\rm eff}}\mu}
{\sqrt{E^1}}\mathcal E^{(\mu)}
\Bigg),
\end{aligned}
\end{equation}
where we define
\begin{equation}\label{eq:definesD}
D=(\mathcal L_{\mathfrak N}E^1)^2\partial_{M_{\rm eff}}\mu
-8\mu N^2\partial_{s_2}^2M_{\rm eff}.
\end{equation}
To express $\heff$ and $H_x$ entirely in terms of $N$, $N^x$, $E^1$, $E^2$, and their derivatives, we first need to eliminate $\partial_{s_2}^2M_{\rm eff}$ and $M_{\rm eff}$ in favor of these variables.  To this end, we return to the covariance equation \eqref{eq:determinMeff}. Taking derivatives with respect to $s_2$ for both sides of Eq. \eqref{eq:determinMeff}, we obtain
\begin{equation}\label{eq:ds22M}
\partial_{s_2}^2M_{\rm eff}=\frac{ \partial_{M_{\rm eff}}\mu}{8 \mu}\left(\frac{\mathcal L_{\mathfrak N}E^1}{N}\right)^2+\frac{1}{8} E^1 \mu \partial_{M_{\rm eff}}\mathcal Z,
\end{equation}
where the first equation in Eq. \eqref{eq:dtE1} is applied. Furthermore, combining the first equation in Eq.~ \eqref{eq:dtE1} and Eq.~\eqref{eq:determinMeff}, we obtain
\begin{equation}\label{eq:getM}
\left(\frac{\mathcal L_{\mathfrak N}E^1}{N}\right)^2=\mu E^1\left(\left(\frac{\partial_xE^1}{E^2}\right)^2+\mathcal Z\right).
\end{equation}
This equation allows us to determine $M_{\rm eff}$ implicitly in terms of $N$, $E^1$, $E^2$, and their derivatives. Substituting the result and Eq.~\eqref{eq:ds22M} into \eqref{eq:HeffHx1} and \eqref{eq:HeffHx2}, we can finally express $\heff$ and $H_x$ entirely in terms of $N$, $E^1$, $E^2$, and their derivatives.

Now let us consider $\{K_2,\mathring H_{\rm eff}[\lambda]+H_x[N^x]\}\Big|_{\lambda=\mathring N}$. Using the Hamiltonian constraint and $\{K_2,H_x[N^x]\}=N^x\partial_xK_2$, we obtain
\begin{equation}\label{eq:dtk21}
\begin{aligned}
&\{K_2,\mathring H_{\rm eff}[\lambda]+H_x[N^x]\}\Big|_{\lambda=\mathring N}\\
=&N^x\partial_xK_2-2 N\mathcal R-2N\partial_{s_1}M_{\rm eff}\\
&-\frac{2NH_x}{(E^2)^2}\partial_{s_2s_4}M_{\rm eff}+\\
&\frac{\partial_xE^1}{E^2}\Bigg(-\frac{2\partial_xN }{\partial_xE^1}+2 N  \partial_{M_{\rm eff}}\mathcal R \Bigg)\partial_{s_4}M_{\rm eff}+\\
&\frac{N\partial_xE^1 \mu_4}{2\mu(E^2)^2} H_{\rm eff}.
\end{aligned}
\end{equation}
Moreover, from the first equation of Eq.~\eqref{eq:dtE1}, we have
\begin{equation}\label{eq:dtk22}
\begin{aligned}
&\partial_tK_2-N^x\partial_xK_2\\
={}&\frac{\mathcal L_{\mathfrak N}^{\,2}E^1-\frac{1}{N}(\mathcal L_{\mathfrak N}E^1)(\partial_tN-N^x\partial_xN)}{2N\partial_{s_2}^2M_{\rm eff}}\\
&+\frac{\partial_xE^1\,\mathcal L_{\mathfrak N}E^2-E^2\partial_x(\mathcal L_{\mathfrak N}E^1)}{(E^2)^2\partial_{s_2}^2M_{\rm eff}}\partial_{s_2s_4}M_{\rm eff}\\
&-\frac{\mathcal L_{\mathfrak N}E^1}{\partial_{s_2}^2M_{\rm eff}}\partial_{s_1s_2}M_{\rm eff}.
\end{aligned}
\end{equation}
Here, we define
\begin{equation*}
\mathcal L_{\mathfrak N}^{\,2}E^1
:=\mathcal L_{\mathfrak N}(\mathcal L_{\mathfrak N}E^1).
\end{equation*}
Combining Eqs.~\eqref{eq:dtk21} and \eqref{eq:dtk22} yields
\begin{equation}\label{eq:poiK1}
\begin{aligned}
&\mathring{\mathscr D}K_2=-\frac{\mathcal L_{\mathfrak N}^{\,2}E^1-\frac{1}{N}(\mathcal L_{\mathfrak N}E^1)(\partial_tN-N^x\partial_xN)}{2N\partial_{s_2}^2M_{\rm eff}}\\
&-2N\partial_{s_1}M_{\rm eff}
+\frac{\mathcal L_{\mathfrak N}E^1}{\partial_{s_2}^2M_{\rm eff}}\partial_{s_1s_2}M_{\rm eff}\\
&-\Bigg[\frac{2NH_x}{(E^2)^2}+\frac{\partial_xE^1\,\mathcal L_{\mathfrak N}E^2-E^2\partial_x(\mathcal L_{\mathfrak N}E^1)}{(E^2)^2\partial_{s_2}^2M_{\rm eff}}\Bigg]\partial_{s_2s_4}M_{\rm eff}\\
&+\frac{\partial_xE^1}{E^2}\Bigg(-\frac{2\partial_xN }{\partial_xE^1}+2 N  \partial_{M_{\rm eff}}\mathcal R \Bigg)\partial_{s_4}M_{\rm eff}\\
&+\frac{N\partial_xE^1 \mu_4}{2\mu(E^2)^2} H_{\rm eff}-2N\mathcal R.
\end{aligned}
\end{equation}
Moreover, using \eqref{eq:dtE2} to solve $H_x$ and substituting the result, we get 
\begin{equation}\label{eq:Hxplus}
\begin{aligned}
&\frac{2NH_x}{(E^2)^2}+\frac{\partial_xE^1\,\mathcal L_{\mathfrak N}E^2-E^2\partial_x(\mathcal L_{\mathfrak N}E^1)}{(E^2)^2\partial_{s_2}^2M_{\rm eff}}\\
=&\frac{\partial_xE^1}{\mu(E^2)^2\partial_{s_2}^2M_{\rm eff}}\Bigg[
\frac{\partial_{M_{\rm eff}}\mu\,\mathcal L_{\mathfrak N}E^1}{4}H_{\rm eff}\\
&+\mu E^2\mathcal L_{\mathfrak N}E^1\,\partial_{M_{\rm eff}}\mathcal R
-\frac{\mu E^2\mathcal L_{\mathfrak N}E^1\,\partial_xN}{N\partial_xE^1}\\
&-\frac{\mu^{3/2}E^2\mathcal L_{\mathfrak N}E^1\,\partial_{M_{\rm eff}}\mu}{2\sqrt{E^1}}\mathcal E^{(\mu)}
\Bigg].
\end{aligned}
\end{equation}
Taking advantage of Eq.~\eqref{eq:Hxplus}, we get 
\begin{equation}\label{eq:poiK1p}
\begin{aligned}
&\mathring{\mathscr D}K_2=-2N\mathcal R
+\frac{1}{8\partial_{s_2}^2M_{\rm eff}}
\Bigg[
\frac{4\mu E^1\partial_xE^1\partial_xN}
{(E^2)^2}\\
&-\frac{4\mu NE^1(\partial_xE^1)^2}
{(E^2)^2}\partial_{M_{\rm eff}}\mathcal R+2\mu NE^1\partial_{s_1}\mathcal Z\\
&-\frac{NE^1(\partial_xE^1)^2}
{(E^2)^3}
\partial_{M_{\rm eff}}\mu\,H_{\rm eff}+\frac{2(\mathcal L_{\mathfrak N}E^1)^2}
{\mu N}\partial_{s_1}\mu\\
&+\frac{4\mathcal L_{\mathfrak N}E^1(\partial_tN-N^x\partial_xN)}{N^2}
-\frac{4\mathcal L_{\mathfrak N}^{\,2}E^1}{N}+\frac{2(\mathcal L_{\mathfrak N}E^1)^2}{NE^1}\Bigg]\\
&+\frac{N\sqrt{\mu}\,\partial_xE^1}
{E^2\sqrt{E^1}}
\left(
\mu_4+
\frac{\mu E^1\partial_xE^1}
{4E^2\partial_{s_2}^2M_{\rm eff}}
\partial_{M_{\rm eff}}\mu
\right)\mathcal E^{(\mu)}.
\end{aligned}
\end{equation}
where need to relations following from the covariance equation \eqref{eq:determinMeff} 
\begin{equation}\label{eq:ds124Meff}
\begin{aligned}
\partial_{s_2s_4}M_{\rm eff}=&\frac{\partial_{s_2}^2M_{\rm eff}}{\partial_{s_2}M_{\rm eff}}\partial_{s_4}M_{\rm eff}+\frac{s_1s_4\mu}{4\partial_{s_2}M_{\rm eff}}\\
\partial_{s_1s_2}M_{\rm eff}=&\frac{\partial_{s_2}^2M_{\rm eff}}{\partial_{s_2}M_{\rm eff}}\partial_{s_1}M_{\rm eff}+\frac{ \mu s_1 \partial_{s_1}\mathcal Z}{8 \partial_{s_2}M_{\rm eff}}\\
&+\frac{\partial_{s_2}M_{\rm eff} \left(s_1 \partial_{s_1}\mu+\mu )\right)}{2\mu s_1}
\end{aligned}
\end{equation} 

Define $\mathring\et_{\mathfrak N\mathfrak N}:=\mathring\et_{\mu\nu}\mathfrak N^\mu\mathfrak N^\nu$, $\mathring\et_{\mathfrak Nx}:=\mathring\et_{\mu\nu}\mathfrak N^\mu \delta_x^\nu$.  Eq.~\eqref{eq:einsteineqs} gives
\begin{equation}
\begin{aligned}
\mathring\et_{\mathfrak N\mathfrak N}
=-\frac{N^2\sqrt{\mu E^1}}{2E^2}H_{\rm eff},\\
\mathring\et_{\mathfrak Nx}
=-\frac{N\sqrt{\mu E^1}}{2E^2}H_x,\\
\mathring\et_{xx}=\frac{(E^2)^2}{2 N\sqrt{\mu E^1}}\mathring{\mathscr D}K_2.
\end{aligned}
\end{equation}
Substituting Eqs.~\eqref{eq:HeffHx1}, \eqref{eq:HeffHx2} and \eqref{eq:poiK1p} we obtain
\begin{equation}\label{eq:finaeinsteinten-tt}
\begin{aligned}
\mathring\et_{\mathfrak N\mathfrak N}
=&-\frac{2\mu N^2\sqrt{\mu E^1}}
{D E^2\partial_xE^1}
\Bigg[
\partial_xE^1\mathcal L_{\mathfrak N}E^1\mathcal L_{\mathfrak N}E^2\\
&+4N^2E^2
\left(\mathcal R\partial_xE^1
+\partial_xM_{\rm eff}\right)
\partial_{s_2}^2M_{\rm eff}\\
&-NE^2\mathcal L_{\mathfrak N}E^1
\partial_x\left(\frac{\mathcal L_{\mathfrak N}E^1}{N}\right)\\
&-E^2\partial_xE^1(\mathcal L_{\mathfrak N}E^1)^2
\partial_{M_{\rm eff}}\mathcal R
\Bigg]\\
&-\frac{\mu^2N^2(\mathcal L_{\mathfrak N}E^1)^2
\partial_{M_{\rm eff}}\mu}{D}
\mathcal E^{(\mu)},
\end{aligned}
\end{equation}
and
\begin{equation}\label{eq:finaeinsteinten-tx}
\begin{aligned}
\mathring\et_{\mathfrak Nx}
=&\frac{N^2\sqrt{\mu E^1}}{D E^2}
\Bigg[
2\mu E^2\mathcal L_{\mathfrak N}E^1\partial_xE^1
\partial_{M_{\rm eff}}\mathcal R\\
&-E^2\mathcal L_{\mathfrak N}E^1
\left(\mathcal R\partial_xE^1
+\partial_xM_{\rm eff}\right)
\partial_{M_{\rm eff}}\mu\\
&-2\mu\partial_xE^1\mathcal L_{\mathfrak N}E^2
+2\mu NE^2
\partial_x\left(\frac{\mathcal L_{\mathfrak N}E^1}{N}\right)
\Bigg]\\
&-\frac{\mu^2 N^2\partial_xE^1\mathcal L_{\mathfrak N}E^1
\partial_{M_{\rm eff}}\mu}{D}
\mathcal E^{(\mu)},
\end{aligned}
\end{equation}
and
\begin{equation}\label{eq:finaeinsteinten-xx}
\begin{aligned}
\mathring\et_{xx}
=&\frac{(E^2)^2}
{16N\sqrt{\mu E^1}
\partial_{s_2}^2M_{\rm eff}}
\Bigg[
\frac{2(\mathcal L_{\mathfrak N}E^1)^2}{NE^1}\\
&+\frac{4\mu E^1\partial_xE^1\partial_xN}
{(E^2)^2}
+2\mu NE^1\partial_{s_1}\mathcal Z\\
&-\frac{4\mu NE^1(\partial_xE^1)^2}
{(E^2)^2}\partial_{M_{\rm eff}}\mathcal R+\frac{2(\mathcal L_{\mathfrak N}E^1)^2}
{\mu N}\partial_{s_1}\mu\\
&+\frac{4\mathcal L_{\mathfrak N}E^1(\partial_tN-N^x\partial_xN)}{N^2}
-\frac{4\mathcal L_{\mathfrak N}^{\,2}E^1}{N}
\Bigg]\\
&-\frac{(E^2)^2}{\sqrt{\mu E^1}}\mathcal R+\frac{(\partial_xE^1)^2\partial_{M_{\rm eff}}\mu}
{8\mu N^2\partial_{s_2}^2M_{\rm eff}}
\mathring\et_{\mathfrak N\mathfrak N}\\
&+\frac{E^2\partial_xE^1\partial_{M_{\rm eff}}\mu}{2E^1}
\left(
\frac{\mu E^1\partial_xE^1}
{4E^2\partial_{s_2}^2M_{\rm eff}}
+\partial_{s_4}M_{\rm eff}
\right)\mathcal E^{(\mu)}.
\end{aligned}
\end{equation}
Here, $D$ is defined in Eq.~\eqref{eq:definesD}, and the Hamiltonian constraint $\heff$ appearing in $\mathring{\mathscr D}K_2$ has been expressed in terms of $\mathring\et_{\mathfrak N\mathfrak N}$.

For a given metric, when $\mathcal E^{(\mu)}=0$, the effective Einstein tensor $\mathring\et_{ab}$ can be calculated without knowing the explicit functional form of $M_{\rm eff}$ in terms of the canonical variables. Instead, we only need Eqs.~\eqref{eq:ds22M} and \eqref{eq:getM}. The former determines $\partial_{s_2}^2M_{\rm eff}$, while the latter implicitly determines $M_{\rm eff}$, both in terms of the metric components and their derivatives.
However, when $\mathcal E^{(\mu)}\neq0$, the explicit functional form of $M_{\rm eff}(s_1,s_2,s_4)$ is required, since $\partial_{s_4}M_{\rm eff}$ appears in the resulting expressions. In this case, once $M_{\rm eff}(s_1,s_2,s_4)$ is known, we can solve Eq.~\eqref{eq:getM} for $s_2=K_2$ in terms of  the metric components and their derivatives, and then substitute the resulting expression for $K_2$ into $\partial_{s_4}M_{\rm eff}$.

%An important observation is that, when $\mathcal E^{(\mu)}\neq0$, the resulting equations generally retain an explicit dependence on $\partial_{s_4}M_{\rm eff}$, which cannot be eliminated. Consequently, the matter-coupled dynamics may depend on the specific choice of the covariant effective Hamiltonian theory, even when different theories admit the same family of vacuum geometries. This contrasts with the case $\mathcal E^{(\mu)}=0$, where these additional terms vanish and the resulting equations can be expressed entirely in terms of the metric variables and the functions characterizing the vacuum geometry. As we show below, minimally coupled null dust belongs to this special class.

\subsection{Explicit expression of $\mathring\et_{\theta\theta}$}\label{sec:einsteinnx01}
The most challenging part of calculating $\mathring\et_{\theta\theta}$ is evaluating $\mathring{\mathscr D}K_1$, including the shift contribution. The key observation used in our calculation is the identity
\begin{equation}
\begin{aligned}
&\{H_x,\mathring H_{\rm eff}[\lambda]+H_x[N^x]\}\\
=&\partial_x\lambda\,\mathring H_{\rm eff}
+N^x\partial_xH_x+2(\partial_xN^x)H_x.
\end{aligned}
\end{equation}
It leads to
\begin{equation}\label{eq:DK1tricy}
\begin{aligned}
\partial_x\mathring N \mathring H_{\rm eff}=&-\frac{1}{2}\partial_xE^1\mathring{\mathscr D}K_1+E^2\partial_x\left(\mathring{\mathscr D}K_2\right)\\
&+\left[\partial_t-N^x\partial_x-2(\partial_xN^x)\right]H_x\\
&+\frac{\sqrt{\mu}\mu_2NE^2\partial_xK_2}{\sqrt{E^1}}\mathcal E^{(\mu)}
\end{aligned}
\end{equation}
where the last two equations in Eq. \eqref{eq:finaltensor} must be applied. 
Using Eq.~\eqref{eq:einsteineqs} to express $\mathring H_{\rm eff}$, $H_x$, and $\mathring{\mathscr D}K_2$ in terms of $\mathring\et_{\mathfrak N\mathfrak N}$, $\mathring\et_{\mathfrak Nx}$, and $\mathring\et_{xx}$, respectively, we obtain from Eq. \eqref{eq:DK1tricy}
\begin{equation}
\begin{aligned}
\mathring{\mathscr D}K_1
={}&\frac{4(\partial_x\mathring N)E^2\mathring\et_{\mathfrak N\mathfrak N}}{N^2\mu\sqrt{E^1}\partial_xE^1}+\frac{4E^2}{\partial_xE^1}\partial_x\left(\frac{N\sqrt{\mu E^1}\mathring\et_{xx}}{(E^2)^2}\right)\\
&-\frac{4}{\partial_xE^1}
\left[\partial_t-N^x\partial_x-2(\partial_xN^x)\right]
\left(\frac{E^2\mathring\et_{\mathfrak Nx}}{N\sqrt{\mu E^1}}\right)\\
&+\frac{2\sqrt{\mu}\mu_2NE^2\partial_xK_2}{\partial_xE^1\sqrt{E^1}}\mathcal E^{(\mu)}.
\end{aligned}
\end{equation}
Substituting this result into $\mathring\et_{\theta\theta}=\mathcal T_{\theta\theta}$, which is equal to
\begin{equation}\label{eq:lasteq1}
\begin{aligned}
\mathring{\mathscr D}K_1+\frac{2 N\sqrt{\mu E^1}}{E^1E^2}\mathring \et_{xx}=\frac{4N\sqrt{\mu}E^2}{\sqrt{E^1}}\mathcal T_{\theta\theta},
\end{aligned}
\end{equation}
we finally have
\begin{equation}\label{eq:final2}
\begin{aligned}
&N\sqrt{\mu}\partial_xE^1  \mathcal T_{\theta\theta}\\
=&-\partial_x\left(\frac{1}{\mathring N}\right)\mathring\et_{\mathfrak N\mathfrak N}+\partial_x\left(\frac{ N\sqrt{\mu} E^1}{(E^2)^2}\mathring  \et_{xx}\right)\\
&-\frac{\sqrt{E^1}}{E^2}
\left[\partial_t-N^x\partial_x-2(\partial_xN^x)\right]
\left(\frac{E^2\mathring\et_{\mathfrak Nx}}{N\sqrt{\mu E^1}}\right)\\
&+\frac{\sqrt{\mu}\mu_2N\partial_xK_2}{2}\mathcal E^{(\mu)}
\end{aligned}
\end{equation}

To simplify this equation, we note that
\begin{equation}
\sqrt{-\mathring g}\mathring g^{ab}
=-\frac{E^2}{N\sqrt{\mu E^1}}\mathfrak N^a\mathfrak N^b
+\frac{N\sqrt{\mu E^1}}{E^2}\delta^a_1\delta^b_1.
\end{equation}
Equation \eqref{eq:final2} can therefore be rewritten as
\begin{equation}\label{eq:second2}
\begin{aligned}
&N\sqrt{\mu}\partial_xE^1\,\mathcal T_{\theta\theta}\\
=&\frac{\sqrt{E^1}}{E^2}\Big[
\partial_a\left(\sqrt{-\mathring g}\,\mathring\et^a{}_{x}\right)
-\sqrt{-\mathring g}\,\mathring\Gamma^a_{xc}\mathring\et^c{}_a
\Big]\\
&+\frac{\sqrt{\mu}\mu_2N\partial_xK_2}{2}\mathcal E^{(\mu)}
\end{aligned}
\end{equation}
where we define 
\begin{equation}
\mathring\et^a{}_c=\mathring g^{ab}\mathring\et_{bc},
\end{equation}
and note that
\begin{equation*}
\begin{aligned}
&\frac{\sqrt{-\mathring g}\sqrt{E^1}}{E^2}\mathring\Gamma^a_{xc}\mathring \et^c{}_a\\
=&\partial_x\left(\frac{1}{\mathring N}\right)\mathring\et_{\mathfrak N\mathfrak N}
-\frac{\partial_xN^x}{\mathring N}\mathring\et_{\mathfrak Nx}-\frac{\mathring N\sqrt{E^1}}{E^2}\partial_x\left(\frac{\sqrt{E^1}}{E^2}\right)\mathring\et_{xx}.
\end{aligned}
\end{equation*}
Using the definition of the covariant divergence associated with
$\mathring g_{ab}$ and
$\sqrt{-\mathring g}=N\sqrt{\mu}E^2/\sqrt{E^1}$,
Eq.~\eqref{eq:second2} reduces to
\begin{equation}\label{eq:lastequation}
\mathring\nabla_a\mathring\et^a{}_x+\frac{\mu_2\partial_xK_2}{2}\mathcal E^{(\mu)}
=\partial_xE^1\,\mathcal T_{\theta\theta}.
\end{equation}
Here, $\mathring\nabla_a\mathring\et^a{}_x$ denotes the $x$-component of $\mathring\nabla_a\mathring\et^a{}_b$. We thus get 
\begin{equation}\label{eq:gthth}
\mathring\et_{\theta\theta}=\frac{1}{\partial_xE^1}\left(\mathring\nabla_a\mathring\et^a{}_x+\frac{\mu_2\partial_xK_2}{2}\mathcal E^{(\mu)}\right).
\end{equation}

\section{Gravitational dynamics with matter in  diagonal areal gauge}\label{sec:IV}
Having established the effective Einstein equation in the preceding section, we now turn to the gravitational dynamics in the presence of matter. To describe general dynamical spherically symmetric geometries, we work in the diagonal areal gauge and promote the constant mass parameter $m$ to a spacetime-dependent function $\mathfrak m(t,x)$. Since a general metric in this gauge contains two independent functions, we introduce an additional function $\beta(t,x)$ and write
\begin{equation}\label{eq:givenmetricgeneral}
\dd s^2=-\beta(t,x)^2F(x,\mathfrak m)\dd t^2+\frac{1}{H(x,\mathfrak m)}\dd x^2+x^2\dd\Omega^2.
\end{equation}
Provided that $H(x,m)$ is locally invertible with respect to $m$, this parametrization describes a general spherically symmetric geometry in the diagonal areal gauge. From the metric, we can read
\begin{equation}\label{eq:componentfrommetric}
\begin{aligned}
N=\beta\sqrt{F},\quad N^x=0,\quad E^1=x^2,\quad \frac{(E^2)^2}{\mu}=\frac{x^2}{H}. 
\end{aligned}
\end{equation}

To calculate $\mathring{\et}_{\mu\nu}$, we first use Eq.~\eqref{eq:getM} to determine $M_{\rm eff}$ in terms of the metric components and their derivatives, as discussed above. Substituting Eq.~\eqref{eq:componentfrommetric} into Eq.~\eqref{eq:getM}, we obtain
\begin{equation}\label{eq:solvMeffvaidya}
0=\frac{x^2}{4}\Big[4H(x,\mathfrak m)+\mu(x^2,M_{\rm eff})\mathcal Z(x^2,M_{\rm eff})\Big].
\end{equation}
To solve this equation for $M_{\rm eff}$, we need to know the functions $\mu(s_1,M_{\rm eff})$ and $\mathcal Z(s_1,M_{\rm eff})$. Recall that these functions were determined by prescribing $M_{\rm eff}=M(x;m)$ in the vacuum sector and requiring the static metric \eqref{eq:metric00} to be a vacuum solution of the theory. 
These requirements led to the reconstruction equations \eqref{eq:reconstructionEqs}. In the vacuum case, Eq.~\eqref{eq:getM} still holds. Substituting the static vacuum solution, for which $N=\sqrt{F(x,m)}$, $E^1=x^2$, and $(E^2)^2/\mu=x^2/H(x,m)$, into this equation, we must therefore recover $M_{\rm eff}=M(x;m)$. In other words, the reconstruction procedure guarantees that
\begin{equation}
0=\frac{x^2}{4}\Big[4H(x,m)+\mu(x^2,M(x;m))\mathcal Z(x^2,M(x;m))\Big],
\end{equation}
which holds  for arbitrary $m$. Comparing this equation with Eq.~\eqref{eq:solvMeffvaidya}, we conclude that the latter admits the solution
\begin{equation}
M_{\rm eff}=M(x,\mathfrak m).
\end{equation}
Consequently, all quantities depending on $s_1$ and $M_{\rm eff}$, including $\mu$, $\mathcal Z$, $\mathcal R$, and their derivatives with respect to $M_{\rm eff}$, retain the same functional forms as in the static vacuum case, with the constant mass parameter $m$ replaced by the spacetime-dependent function $\mathfrak m$. In particular, 
we get from the reconstruction equation \eqref{eq:reconstructionEqs} that 
\begin{equation}
\mu=\frac{H}{F(\partial_mM)^2},
\end{equation}
which, together with Eq. \eqref{eq:componentfrommetric}, leads to 
\begin{equation}\label{eq:E2inF}
E^2=\frac{x}{\sqrt F\,\partial_mM}.
\end{equation}
Furthermore, the above arguments also apply to $\partial_{s_2}^2M_{\rm eff}$, which is determined by Eq.~\eqref{eq:ds22M}. With this result, we can now calculate $\mathring{\et}_{\mu\nu}$. 

We first calculate $\mathring\et_{tt}$. Since $\partial_tE^1=0$, Eq.~\eqref{eq:definesD} simplifies to
\begin{equation}\label{eq:definesDppp}
D=-8\mu N^2\partial_{s_2}^2M_{\rm eff}.
\end{equation}
Using this result, we could get from Eq.~\eqref{eq:finaeinsteinten-tt} for $\mathring\et_{tt}$ as
\begin{equation}
\mathring\et_{tt}
=\frac{N^2\sqrt{\mu E^1}}{\partial_xE^1}
\left(\mathcal R\partial_xE^1+\partial_xM_{\rm eff}\right),
\end{equation}
where we have assumed regular matter, such that the scalar $\mathcal E^{(\mu)}$, which is proportional to the trace of the two-dimensional energy-momentum tensor, remains finite.
Substituting $E^1=x^2$ and $M_{\rm eff}=M(x;\mathfrak m)$, we obtain
\begin{equation}
\begin{aligned}
\partial_xM_{\rm eff}
=&\left.\partial_xM(x;m)\right|_{m=\mathfrak m}+\partial_mM(x;\mathfrak m)\partial_x\mathfrak m.
\end{aligned}
\end{equation}
Then, we get 
\begin{equation}
\begin{aligned}
\mathring\et_{tt}
=&\frac{N^2\sqrt{\mu E^1}}{\partial_xE^1}
\left(\mathcal R\partial_xE^1+\left.\partial_xM(x;m)\right|_{m=\mathfrak m}\right)\\
&+\frac{N^2\sqrt{\mu E^1}}{\partial_xE^1}\partial_mM\partial_x\mathfrak m.
\end{aligned}
\end{equation}
The first term is exactly $\mathring\et_{tt}$ evaluated on the static vacuum solution, with $m$ replaced by $\mathfrak m$, and therefore vanishes. To simplify the second term, we note that from the second relation of Eqs. \eqref{eq:reconstructionEqs} and \eqref{eq:componentfrommetric}
\begin{equation}
\begin{aligned}
N^2\sqrt{\mu}(\partial_m M)=\beta^2\sqrt{FH},
\end{aligned}
\end{equation}
where we assume that $\partial_mM>0$ consistent with the convention adopted in Eq.~\eqref{eq:NoindmM}. 
These results thus lead to 
\begin{equation}\label{eq:Gtt}
\begin{aligned}
\mathring\et_{tt}=\frac{\beta^2 \sqrt{FH} }{2}\partial_x\mathfrak m.
\end{aligned}
\end{equation}

%This example illustrates that the key step in the calculation is to separate the terms involving derivatives of $\mathfrak m$, such as $\partial_x\mathfrak m$ and $\partial_t\mathfrak m$, from those independent of these derivatives. The latter is identical to the corresponding expression in the static vacuum case, with the constant mass parameter $m$ replaced by $\mathfrak m$, and therefore vanishes. We therefore only need to consider the terms involving $\partial_x\mathfrak m$ and $\partial_t\mathfrak m$.
%The same strategy can be applied to calculate the other components of $\mathring\et$. However, one must take into account the additional factor $\beta$ in $N=\beta\sqrt F$, whose derivatives also contribute to the resulting expressions.

For $\mathring\et_{tx}$, from \eqref{eq:finaeinsteinten-tx}, we could get 
\begin{equation}\label{eq:gtx}
\begin{aligned}
\mathring\et_{tx}=\frac{\sqrt{\mu E^1}\partial_xE^1}{4\partial_{s_2}^2M_{\rm eff}}\frac{\partial_tE^2}{E^2}.
\end{aligned}
\end{equation}
According to Eq. \eqref{eq:E2inF}, we get 
\begin{equation}\label{eq:dE2}
\begin{aligned}
\frac{\partial_t E^2}{E^2}=\frac{\partial_{M_{\rm eff}}\mathcal Z}{8\partial_mM F} \partial_t\mathfrak m
\end{aligned}
\end{equation}
where we used that, according to the last relation in Eq. \eqref{eq:reconstructionEqs},  
\begin{equation}\label{eq:dMZ}
\begin{aligned}
\partial_{M_{\rm eff}}\mathcal Z=-4\left(\partial_mF\partial_mM+2F\partial_m^2M\right)
\end{aligned}
\end{equation}
Moreover, according to Eq. \eqref{eq:ds22M}, we get
\begin{equation}\label{eq:ds22Mp}
\partial_{s_2}^2M_{\rm eff}=\frac{1}{8} E^1 \mu \partial_{M_{\rm eff}}\mathcal Z.
\end{equation}
Substituting Eqs. \eqref{eq:dE2} and \eqref{eq:ds22Mp}, as well as $E^1=x^2$ and the second relation in Eq. \eqref{eq:reconstructionEqs}, into Eq. \eqref{eq:gtx}, we finally get
\begin{equation}\label{eq:Gtx}
\mathring\et_{tx}=\frac{1}{2\sqrt{FH}} \partial_t\mathfrak m
\end{equation}

Finally, we evaluate $\mathring\et_{xx}$ using Eq.~\eqref{eq:finaeinsteinten-xx}. To simplify the calculation, we first establish the following useful identity:
\begin{equation}\label{eq:toprove}
\frac{\mu E^1\partial_xE^1}
{4E^2\partial_{s_2}^2M_{\rm eff}}
+\partial_{s_4}M_{\rm eff}=0.
\end{equation}
To begin with, the first equation in Eq.~\eqref{eq:ds124Meff} yields
\begin{equation}\label{eq:Meff1}
\frac{\mu E^1\partial_xE^1}
{4E^2\partial_{s_2}^2M_{\rm eff}}
+\partial_{s_4}M_{\rm eff}=\frac{\partial_{s_2}M_{\rm eff}\partial_{s_2}\partial_{s_4}M_{\rm eff}}{\partial_{s_2}^2M_{\rm eff}}
\end{equation}
On the other hand, evaluating the second equation in Eq.~\eqref{eq:finaltensor}, which is equivalent to Eq.~\eqref{eq:dtE1}, on the metric~\eqref{eq:givenmetricgeneral}, and noting that $\partial_tE^1=0$, we obtain
\begin{equation}\label{eq:ds2m}
\partial_{s_2}M_{\rm eff}=0.
\end{equation}
Then, Eq. \eqref{eq:toprove} is just a consequence of substituting this result into \eqref{eq:Meff1}. 

Using Eq.~\eqref{eq:toprove}, we can simplify Eq. \eqref{eq:finaeinsteinten-xx} to 
\begin{equation}\label{eq:finaeinsteinten-xx1}
\begin{aligned}
\mathring\et_{xx}
=&\frac{(E^2)^2}
{16N\sqrt{\mu E^1}
\partial_{s_2}^2M_{\rm eff}}
\Bigg[\frac{4\mu E^1\partial_xE^1\partial_xN}
{(E^2)^2}\\
&+2\mu NE^1\partial_{s_1}\mathcal Z-\frac{4\mu NE^1(\partial_xE^1)^2}
{(E^2)^2}
\partial_{M_{\rm eff}}\mathcal R
\Bigg]\\
&-\frac{(E^2)^2}{\sqrt{\mu E^1}}\mathcal R+\frac{(\partial_xE^1)^2\partial_{M_{\rm eff}}\mu}
{8\mu N^2\partial_{s_2}^2M_{\rm eff}}
\mathring\et_{tt}
\end{aligned}
\end{equation}
To simplify this expression, let us first consider $\partial_{s_1}\mathcal Z$. According to the last equation in Eq. \eqref{eq:reconstructionEqs}, we get
\begin{equation}\label{eq:ds1Z}
\begin{aligned}
&2x\partial_{s_1}\mathcal Z+\partial_{M_{\rm eff}}\mathcal Z\partial_xM\big|_{m=\mathfrak m} \\
=&-4\Big((\partial_mM)^2\partial_xF\big|_{m=\mathfrak m}+2F\partial_mM\partial_{x}\partial_mM\big|_{m=\mathfrak m}\Big).
\end{aligned}
\end{equation}
Here, $\partial_xO\big|_{m=\mathfrak m}$ indicates that the derivative $\partial_x$ is first evaluated with $m$ held fixed, after which $m$ is replaced by the spacetime-dependent function $\mathfrak m(t,x)$. Thus, $\partial_x$ acts only on the explicit $x$ dependence and not on $\mathfrak m$. Moreover, by the first equation in Eq. \eqref{eq:reconstructionEqs}, we have
\begin{equation}\label{eq:dxmM}
\begin{aligned}
\partial_x \partial_m M\big|_{m=\mathfrak m} =-2x\partial_{M_{\rm eff}}\mathcal R\partial_mM
\end{aligned}
\end{equation}
Then, substituting Eqs. \eqref{eq:ds22Mp} and \eqref{eq:dxmM} into Eq. \eqref{eq:ds1Z} yields 
\begin{equation}
\begin{aligned}
&\mu E^1\left(\partial_{s_1}\mathcal Z-\frac{2(\partial_xE^1)^2}{(E^2)^2}\partial_{M_{\rm  eff}}\mathcal R\right)\\
=&-\frac{2xH \partial_xF|_{m=\mathfrak m}}{F}+8\partial_{s_2}^2M_{\rm eff}\mathcal R.
\end{aligned}
\end{equation}
where we have applied Eqs. \eqref{eq:componentfrommetric} and  \eqref{eq:E2inF}, and the first and second relations in Eq. \eqref{eq:reconstructionEqs}.
With this result, we have
\begin{equation}\label{eq:finaeinsteinten-xx11}
\begin{aligned}
\mathring\et_{xx}
=&\frac{(E^2)^2}
{16N\sqrt{\mu E^1}
\partial_{s_2}^2M_{\rm eff}}
\Bigg[\frac{4\mu E^1\partial_xE^1\partial_xN}
{(E^2)^2}\\
&-\frac{4N xH \partial_xF|_{m=\mathfrak m}}{F}\Bigg]+\frac{(\partial_xE^1)^2\partial_{M_{\rm eff}}\mu}
{8\mu N^2\partial_{s_2}^2M_{\rm eff}}
\mathring\et_{tt}
\end{aligned}
\end{equation}
To simplify the term involving $\partial_{M_{\rm eff}}\mu$, we again use the relation given in the reconstruction equation \eqref{eq:reconstructionEqs} to get  
\begin{equation}\label{eq:dmmu}
\begin{aligned}
\partial_{M_{\rm eff}}\mu\,\partial_m M
=
\mu\left[
\frac{\partial_mH}{H}
-\frac{\partial_mF}{F}
-2\frac{\partial_m^2M}{\partial_mM}
\right].
\end{aligned}
\end{equation} 
Moreover, the combination of Eqs. \eqref{eq:componentfrommetric} and  \eqref{eq:E2inF}, we get
\begin{equation}\label{eq:ds2M}
\begin{aligned}
\frac{1}{\partial_{s_2}^2M_{\rm eff}}=-\frac{1}{\mu E^1 \sqrt{F}\partial_m\left(\sqrt{F}\partial_mM\right)}
\end{aligned}
\end{equation}
Substituting Eqs. \eqref{eq:dmmu}, \eqref{eq:ds2M}, \eqref{eq:componentfrommetric}, \eqref{eq:E2inF} and \eqref{eq:Gtt} into Eq. \eqref{eq:finaeinsteinten-xx11}, we get
\begin{equation}\label{eq:finaeinsteinten-xx111}
\begin{aligned}
&\mathring\et_{xx}
=\frac{\partial_x\mathfrak m }
{2\sqrt{FH}}-\frac{\partial_mM \Xi}
{2  \sqrt{H}\partial_m\left(\sqrt{F}\partial_mM\right)}
\end{aligned}
\end{equation}
where we introduced 
\begin{equation}
\Xi=
\frac{\beta_x}{\beta}
+
\partial_m\ln\sqrt{FH}\partial_x\mathfrak m
\end{equation}
for abbreviation.

To calculate $\mathring\et_{\theta\theta}$, we note that $\mu_2=\partial_{M_{\rm eff}}\mu\partial_{s_2}M_{\rm eff}=0$ due to Eq. \eqref{eq:ds2m}. Thus, we have
\begin{equation}
\mathring\et_{\theta\theta}=\frac{1}{2x}\mathring\nabla_a\mathring\et^a{}_x
\end{equation}
leading to 
\begin{equation}\label{eq:Gthth}
\begin{aligned}
\mathring\et_{\theta\theta}={}&
\frac{(\partial_mM)^2}{4x\beta}\sqrt{\frac FH}
\Bigg[
\frac{\partial_x
\left(
\beta^2\sqrt{FH}\partial_x\mathfrak m
\right)}{\beta\sqrt{FH}}-
\partial_t
\left(
\frac{\partial_t\mathfrak m}{\beta FH}
\right)
\Bigg]
\\
&-
\frac{\partial_mM}{4\beta x\sqrt H}
\partial_x
\left[
\beta
\frac{F(\partial_mM)^2}
{\partial_m(\sqrt F\,\partial_mM)}
\,\Xi
\right].
\end{aligned}
\end{equation}

\section{Generalized Vaidya solution}\label{sec:IV5}

In this section, we construct generalized Vaidya solutions for the case $\mathcal E^{(\mu)}=0$. We emphasize that this condition concerns the matter sector rather than imposing an additional restriction on the gravitational theory. According to Eq.~\eqref{eq:TfromTo}, this condition implies that the trace of the four-dimensional energy-momentum tensor is entirely determined by the tangential pressure. A simple example satisfying this condition is null dust, whose energy-momentum tensor has a vanishing trace and no tangential pressure. Nevertheless, the following construction does not assume null dust as the matter source.

In general relativity, the Vaidya solution can be obtained by promoting the constant parameter $m$ of the Schwarzschild metric to a function of a null coordinate. In a modified theory of gravity, however, such a promotion does not automatically yield a solution sourced by standard null dust. The resulting matter source depends on the modified gravitational dynamics and must therefore be determined from the corresponding field equations. Our aim in this section is to show that, within the reconstructed covariant effective theory, this procedure remarkably leads to precisely the standard Vaidya energy-momentum tensor.

Starting from a general spherically symmetric static metric \eqref{eq:metric00}, we define
\begin{equation}
\dd v_\pm=\dd t\pm \frac{1}{\sqrt{FH}}\dd x.
\end{equation}
In each coordinate patch, we may choose either $(v_+,x)$ or $(v_-,x)$ as the coordinates. The metric then takes the form
\begin{equation}
\begin{aligned}
\dd s^2=&-F(x,m)\dd v_\pm^2\pm 2\sqrt{\alpha}\dd v_\pm\dd x+x^2\dd\Omega^2,
\end{aligned}
\end{equation}
where we define 
\begin{equation}
\alpha=\frac{F}{H}. 
\end{equation}
The Vaidya extension of this metric is obtained by promoting the constant mass parameter $m$ to a function of the null coordinate $v_\pm$. The resulting metric takes the form
\begin{equation}\label{eq:vaidyametricextension}
\begin{aligned}
\dd s^2=&-F(x,\mathfrak m(v_\pm))\dd v_\pm^2\pm 2\sqrt{\alpha(x,\mathfrak m(v_\pm)}\dd v_{\pm}\dd x\\
&+x^2\dd\Omega^2.
\end{aligned}
\end{equation}

To apply the results derived in Sec.~\ref{sec:IV}, we need to come back the coordinates corresponding to the diagonal areal gauge. The time coordinate $t$ in this gauge is related to $v_\pm$ by
\begin{equation}
\dd v_\pm=\beta\dd t\pm\frac{1}{\sqrt{FH}}\dd x
\end{equation}
where, unlike the static case with constant $m$, the factor $\beta$ is necessary to ensure the right hand side is an exact differential.  $\beta$ thus satisfies 
\begin{equation}\label{eq:dtaubeta}
\partial_x\beta=\pm \partial_t\left(\frac{1}{\sqrt{FH}}\right)
\end{equation}
With this equation, the metric in $(t,x)$ coordinates returns to the form given in Eq. \eqref{eq:givenmetricgeneral} explicitly with $\mathfrak m=\mathfrak m(v_\pm)$. Noting that $F$ and $H$ depends on $t$ through $\mathfrak m$, we get from Eq. \eqref{eq:dtaubeta} that 
\begin{equation}
\begin{aligned}
\Xi=\frac{\partial_x\beta}{\beta}+\partial_m\ln(\sqrt{FH})\partial_x\mathfrak m =0
\end{aligned}
\end{equation}
Substituting this results into Eq. \eqref{eq:finaeinsteinten-xx111},
we get 
\begin{equation}
\mathring\et_{xx}=\frac{\partial_x\mathfrak m}{2\sqrt{FH}}. 
\end{equation} 

To calculate $\mathring \et_{\theta\theta}$, we note that, in this case,
\begin{equation}
\pm \beta\sqrt{FH}\partial_x\mathfrak m=\partial_t\mathfrak m. 
\end{equation}
Thus, $\mathring\et_{\theta\theta}$ in this case can be simplified to, with considering $\Xi=0$,
\begin{equation}\label{eq:Gthth1}
\begin{aligned}
\mathring\et_{\theta\theta}=&
\frac{\pm (\partial_mM)^2}{4x\beta}\sqrt{\frac FH}
\Bigg[
\frac{\partial_x
\left(
\beta\partial_t\mathfrak m
\right)}{\beta\sqrt{FH}}-
\partial_t
\left(
\frac{\partial_x\mathfrak m}{\sqrt{FH}}
\right)
\Bigg]\\
=&
\frac{(\partial_mM)^2\partial_x\mathfrak m}{4x\beta}\sqrt{\frac FH}
\Bigg[
\pm\partial_x
\beta-
\partial_t
\left(
\frac{1}{\sqrt{FH}}
\right)
\Bigg]\\
=&0.
\end{aligned}
\end{equation}

Combining all of the above results, including the expression of $\mathring \et_{tt}$ and $\mathring \et_{tx}$ given in \eqref{eq:Gtt} and \eqref{eq:Gtx} respectively, we finally get 
\begin{equation}
\begin{aligned}
\mathcal T_{\mu\nu}\dd x^\mu\dd x^\mu=\mathring\et_{\mu\nu}\dd x^\mu\dd x^\nu=\pm\frac12\frac{\dd\mathfrak m}{\dd v_\pm}
\dd v_\pm^2.
\end{aligned}
\end{equation}

Since $\mathcal E^{(\mu)}=0$, the definition of $\mathcal T_{\mu\nu}$ in Eq.~\eqref{eq:calT} implies that $\mathring T_{ab}=\mathcal T_{ab}$ and $T^{(\theta)}=-2\mathcal T_{\theta\theta}=0$.  Using Eq.~\eqref{eq:TfromTo}, we therefore obtain the four-dimensional energy-momentum tensor
\begin{equation}
T_{\mu\nu}\dd x^\mu\dd x^\nu
=\pm\frac{1}{4\pi x^2}
\frac{\dd\mathfrak m}{\dd v_\pm}\dd v_\pm^2.
\end{equation}
This is precisely the standard energy-momentum tensor of null dust, demonstrating that the generalized Vaidya metric constructed above provides a solution to the coupled system of gravity and null dust within our effective framework. Thus, standard null-dust sources can support dynamical black-hole geometries beyond general relativity, including regular black holes. Moreover, since the energy-momentum tensor of ingoing null dust is proportional to $\mathfrak m'(v_+)/x^2$, Hawking-like mass loss, characterized by $\mathfrak m'(v_+)<0$, necessarily entails a violation of the null energy condition. This suggests that quantum gravity corrections to the gravitational dynamics alone may be insufficient to account for Hawking-like evaporation.

\mysection{Conclusion}\label{sec:conclusion}
In this work, building upon the covariant Hamiltonian gravity in spherical symmetry, we have derived the effective Einstein tensor $\mathring{\et}_{\mu\nu}$ associated with a prescribed one-parameter family of static, spherically symmetric metrics. This allows the field equations of the framework to be expressed in an Einstein-like form, $\mathring{\et}_{\mu\nu}=\mathcal T_{\mu\nu}$, where $\mathcal T_{\mu\nu}$ is the tensor related to the energy-momentum tensor of the matter source. The association is understood in the sense that $\mathring{\et}_{\mu\nu}$ vanishes when evaluated on any metric in the prescribed family, demonstrating that these metrics are vacuum solutions of the reconstructed effective theory.

As an application, we construct generalized Vaidya solutions of the reconstructed effective theory. More precisely, starting from a given one-parameter family of static, spherically symmetric metrics characterized by a mass parameter $m$, we first transform the metrics into Eddington–Finkelstein coordinates and then promote $m$ to a function $\mathfrak m(v)$ of the null coordinate $v$. Remarkably, evaluating the effective Einstein tensor on the resulting metric yields a matter source whose energy-momentum tensor takes precisely the standard Vaidya form. This demonstrates that null dust can serve as the matter source for the dynamical extensions of the prescribed static geometries, including regular black holes.  An immediate consequence of the above result is that Hawking-like mass loss in the ingoing Vaidya solution, characterized by $\mathfrak m'(v)<0$, necessarily requires a violation of the null energy condition. This suggests that quantum gravity corrections to the gravitational dynamics alone may be insufficient to account for Hawking-like evaporation within our effective framework.

\begin{acknowledgements}
This work was supported in part by the National Key Research and Development Program of China Grant No. 2021YFC2203001 and in part by the NSFC (Grants No. 12275022, No.~12475046, and No. 12505055) and  ``the Fundamental Research Funds for the Central Universities''.
\end{acknowledgements}
\bibliography{reference}

\end{document}